\documentclass{article}
\usepackage{spconf,amsmath,graphicx,hyperref,cite}
\usepackage{booktabs}
\usepackage{multirow,balance}
\usepackage{subcaption}
\usepackage{array}
\usepackage{siunitx}
\usepackage{soul, xcolor}

\title{Environmental Sound Deepfake Detection Challenge: An Overview}
\name{Han Yin$^{1}$, Yang Xiao$^{2}$, Rohan Kumar Das$^{3}$, Jisheng Bai$^{4,5}$, Ting Dang$^{2}$}
\address{$^{1}$School of Electrical Engineering, KAIST, Daejeon, Republic of Korea\\
$^{2}$University of Melbourne, Australia, $^{3}$Fortemedia Singapore, Singapore,\\
$^{4}$Xi’an University of Posts \& Telecommunications, Xi’an, China,\\
$^{5}$Xi'an Lianfeng Acoustic Technologies Co., Ltd., China}
\begin{document}
\ninept
\maketitle
\begin{abstract}
Recent progress in audio generation models has made it possible to create highly realistic and immersive soundscapes, which are now widely used in film and virtual-reality-related applications. However, these audio generators also raise concerns about potential misuse, such as producing deceptive audio for fabricated videos or spreading misleading information. Therefore, it is essential to develop effective methods for detecting fake environmental sounds. Existing datasets for environmental sound deepfake detection (ESDD) remain limited in both scale and the diversity of sound categories they cover. To address this gap, we introduced EnvSDD, the first large-scale curated dataset designed for ESDD. Based on EnvSDD, we launched the ESDD Challenge, recognized as one of the ICASSP 2026 Grand Challenges. This paper presents an overview of the ESDD Challenge, including a detailed analysis of the challenge results.

\end{abstract}
\section{Introduction}
\label{sec:intro}

Recent progress in audio generation models has made it easy to generate highly realistic environmental sounds, such as alarms, footsteps, and sirens \cite{audioldm}. While such deepfake environmental sounds can be applied for creative audio applications in film, gaming, and virtual reality, they also introduce serious risks when misused for misleading the public or inducing panic. In addition, environmental sound deepfake detection differs from speech anti-spoofing~\cite{asvspoof2021,asvspoof5,aasist,xlsrmamba,xuan25_spsc,rawtf} because it spans more sound types and shows more varied generation artifacts. Therefore, developing effective methods for detecting deepfake environmental sounds has become increasingly important.


To address these gaps, we introduce EnvSDD\footnote{EnvSDD Dataset: https://envsdd.github.io/}~\cite{envsdd}, a large-scale curated dataset for environmental sound deepfake detection that includes both real recordings and generated samples at substantial duration, comprising 45.25 hours of real sound and 316.7 hours of fake sound. Based on EnvSDD, we organized the ICASSP 2026 Environmental Sound Deepfake Detection Challenge (ESDD)\footnote{ESDD Challenge: https://sites.google.com/view/esdd-challenge}$^{,}$\footnote{Challenge Baseline: https://github.com/apple-yinhan/EnvSDD} with two tracks: Track 1 evaluates robustness to unseen audio generators, and Track 2 targets a black-box, low-resource setting where participants must detect deepfakes without prior access to the specific generation frameworks used in evaluation. Beyond reporting leaderboard results, we summarize common effective design choices observed in top systems
in this paper.

\begin{figure}[t]
    \centering
    \begin{subfigure}[t]{\linewidth}
        \centering
        \includegraphics[width=0.8\linewidth]{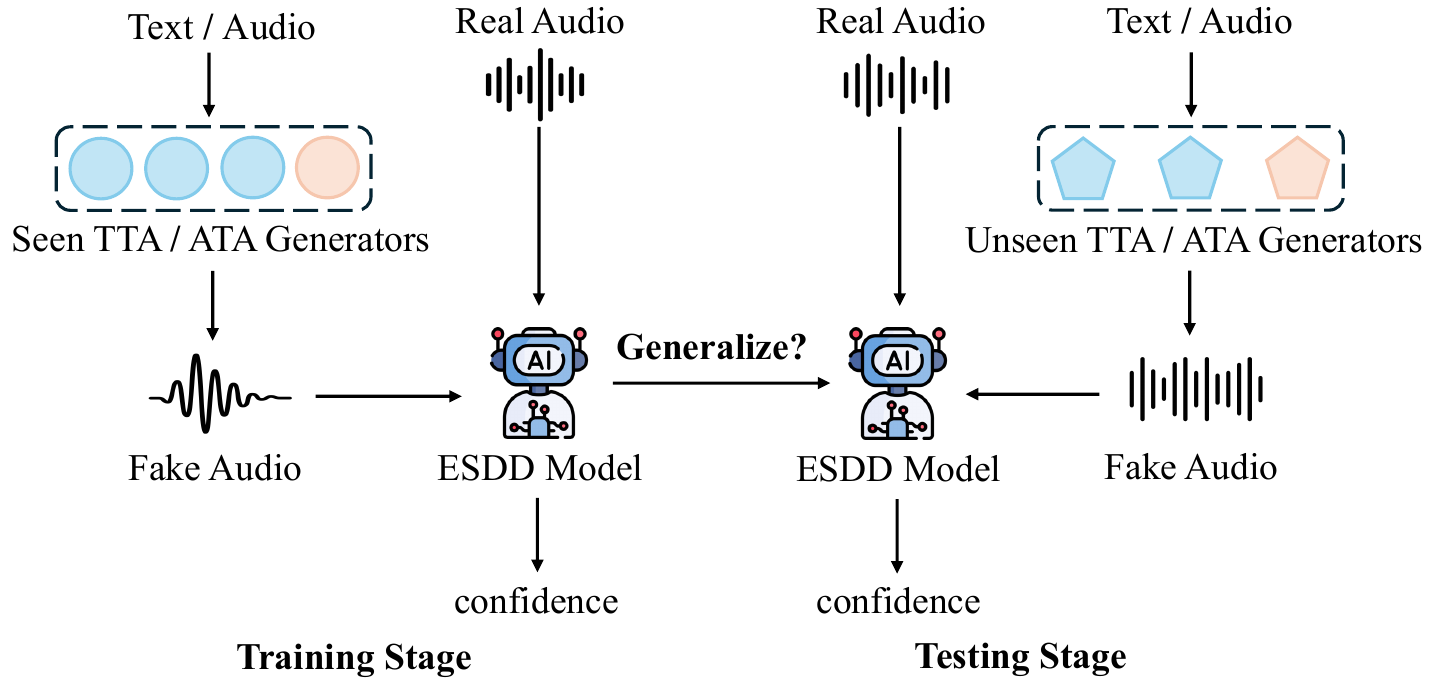}
        \caption{Track 1: ESDD in Unseen Generators.}
        \label{fig:track1}
    \end{subfigure}

    \vspace{2mm} 
    \begin{subfigure}[t]{\linewidth}
        \centering
        \includegraphics[width=0.8\linewidth]{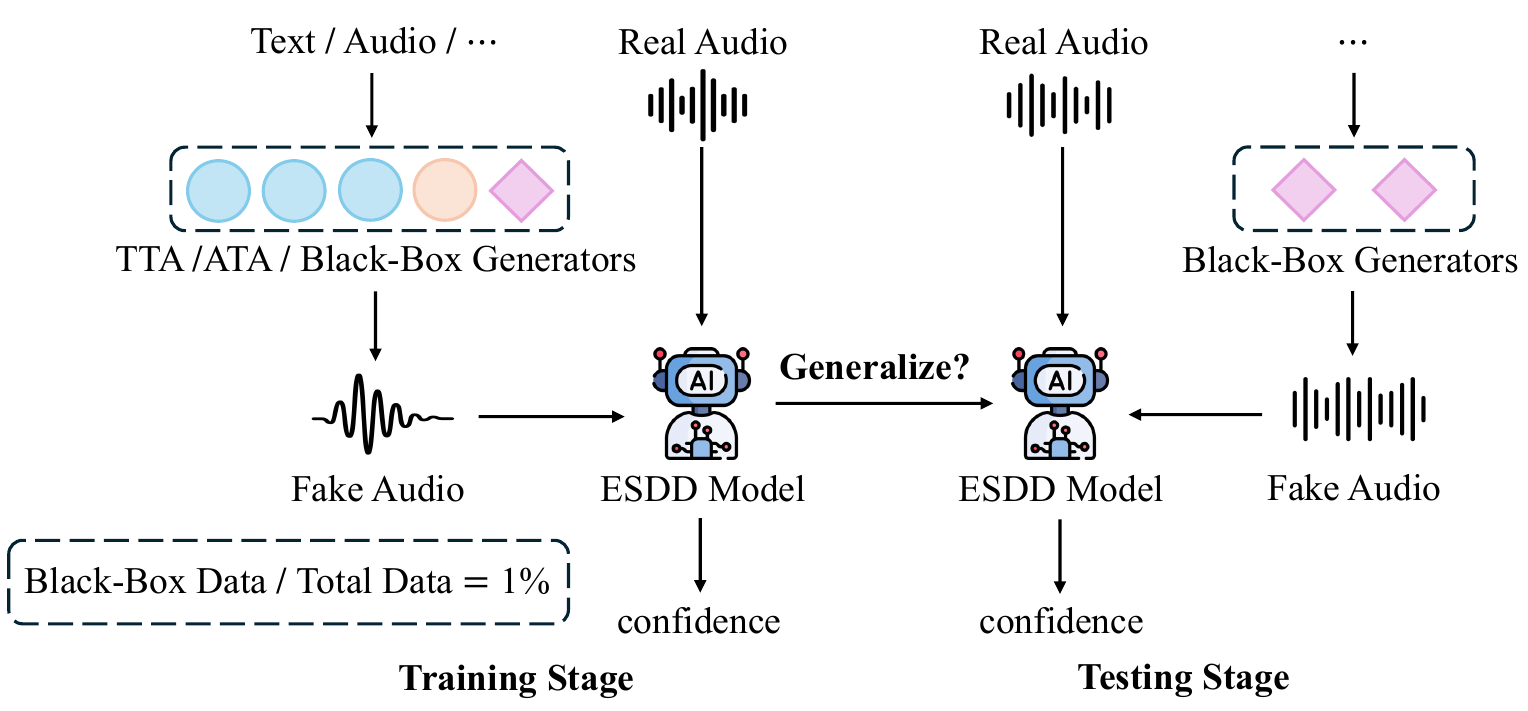}
        \caption{Track 2: Black-Box Low-Resource ESDD.}
        \label{fig:track2}
    \end{subfigure}
\vspace{-3mm}
    \caption{Overview of the two challenge tracks.}
    \label{fig:tracks}
    \vspace{-5mm}
\end{figure}



\section{ESDD Challenge}

As shown in Fig.~\ref{fig:tracks}, we proposed two different tracks: ESDD in Unseen Generators (Track 1) and Black-Box Low-Resource ESDD (Track 2). For more details, please refer to our evaluation plan~\cite{esdd_eval}.

\textbf{ESDD in Unseen Generators:} The primary target of this track is to develop detection models that can effectively identify deepfake environmental sound clips generated by unseen Text-to-Audio (TTA) and Audio-to-Audio (ATA) systems. 
For Track 1, the training and validation splits use generators G01–G04, while the evaluation and test splits use unseen generators G05–G07 from~\cite{esdd_eval}. The evaluation set is a randomly sampled subset of the test set for the progress phase, and the final ranking is based on the test set. Table \ref{tab:stats_tracks_merged} summarizes the number of real and fake clips in each split for Track 1.

\textbf{Black-Box Low-Resource ESDD:} The term ``black-box'' indicates we have no prior knowledge of the specific generation methods used in testing, which may include any generative paradigms beyond TTA and ATA. The ``low-resource'' setting refers to the amount of available black-box training data being severely limited, comprising only 1\% of the total training data. Specifically, for the test set, we use Video-to-Audio models (i.e., FoleyCafter \cite{foleycrafter} and Diff-foley \cite{diff-foley}) for generating the fake data. Notably, this information is not disclosed to participants during the entire competition period. Table \ref{tab:stats_tracks_merged} right reports the split statistics for Track 2.

\begin{table}[t]
\centering
\caption{Statistics of the dataset in Track 1 and Track 2.}
\vspace{-2mm}
\label{tab:stats_tracks_merged}
\renewcommand{\arraystretch}{1.05}
\setlength{\tabcolsep}{5pt}
\small
\resizebox{\columnwidth}{!}{
\begin{tabular}{@{}l
S[table-format=6] S[table-format=6] S[table-format=6]
S[table-format=4] S[table-format=4] S[table-format=5]@{}}
\toprule
& \multicolumn{3}{c}{\textbf{Track 1 dataset}} & \multicolumn{3}{c}{\textbf{Track 2 black-box data}} \\
\cmidrule(lr){2-4}\cmidrule(lr){5-7}
\textbf{Usage}
& {\textbf{\# Real clips}} & {\textbf{\# Fake clips}} & {\textbf{\# Total}}
& {\textbf{\# Real clips}} & {\textbf{\# Fake clips}} & {\textbf{\# Total}} \\
\midrule
Training    & 27811 & 111244 & 139055 & 270  & 1083 & 1353 \\
Validation  & 7942  & 31768  & 39710  & 90   & 361  & 451  \\
Evaluation  & 478   & 1522   & 2000   & 973 & 4014 & 4987 \\
Test        & 1000  & 3000   & 4000   & 1994 & 7980 & 9974 \\
\bottomrule
\end{tabular}}
\vspace{-4mm}
\end{table}

\textbf{Evaluation Metric:} We adopt the equal error rate (EER) to evaluate ESDD performance. Participants submit confidence scores for each sound clip in TXT format, indicating the likelihood that a clip is real. By adjusting the decision threshold, a trade-off is established between the false acceptance rate and the false rejection rate, with the EER occurring at the point where the two rates are equal. A lower EER indicates better detection performance.

\textbf{Baselines:} Following EnvSDD, we develop two systems as the baselines (AASIST and BEATs+AASIST). AASIST \cite{aasist} is an end-to-end system that uses a novel heterogeneou stacking graph attention machanism to learn acoustic features, which has been applied in various speech and singing voice deepfake detection challenges. Building on AASIST, BEATs+AASIST incorporates BEATs \cite{beats} as the front-end, to extract high-level acoustic representations, achieving better performance than AASIST on EnvSDD dataset.

\textbf{Participants:} In total, 97 teams worldwide registered for the challenge. A total of 1,359 and 389 valid submissions were received for Track 1 and Track 2, respectively. Among the participating teams, the majority (82.4\%) were affiliated with academic institutions, while 9.3\% belonged to industry, and the remaining teams reported mixed academic–industrial affiliations. Furthermore, more than half of the participants (56.5\%) indicated prior research experience in the detection of deepfake audio.

\section{Results and Discussions}


We received 97 registrations. Submissions missing the required metadata were excluded from the final leaderboard. Tables~\ref{tab:track_1} and~\ref{tab:track_2} present the final rankings for Track~1 and Track~2, respectively. The evaluation set is randomly sampled as a subset of the test set, and the final ranking is based on the test set. ``Ensem'' denotes the number of models in an ensemble (``1'' indicates a single model).

\begin{table}[t]
\centering
\caption{Top-5 system results of Track 1.}
\vspace{-3mm}
\label{tab:track_1}
\renewcommand{\arraystretch}{1.06}
\setlength{\tabcolsep}{3.6pt}
\small
\resizebox{0.75\linewidth}{!}{%
\begin{tabular}{@{}l| l |c| c c@{}}
\toprule
\multirow{2}{*}{Team} & \multirow{2}{*}{System} & \multirow{2}{*}{Ensem} & \multicolumn{2}{c}{EER (\%)} \\
& & & Eval & Test \\
\midrule
AHU & EAT+AASIST & 5 & -- & \textbf{0.30} \\
DFKI & EAT-L+BiCrossMamba & 4 & \textbf{0.44} & 0.80 \\
CUC\cite{CUC} & SSLAM+FFN & 1 & 1.05 & 1.20 \\
CAU\cite{cau} & BEAT2AASIST & 2 & -- & 1.60 \\
BIT & EAT+ArcFace & 1 & 2.67 & 2.40 \\
\midrule
Baseline 1\cite{esdd_eval} & BEATs+AASIST & 1 & 14.21 & 13.20 \\
Baseline 2\cite{esdd_eval} & AASIST & 1 & 15.26 & 15.02 \\
\bottomrule
\end{tabular}%
}
\vspace{-6mm}
\end{table}

\begin{table}[t]
\centering
\caption{Top-5 system results of Track 2.}
\vspace{-2mm}
\label{tab:track_2}
\renewcommand{\arraystretch}{1.06}
\setlength{\tabcolsep}{3.6pt}
\small
\resizebox{0.8\linewidth}{!}{%
\begin{tabular}{@{}l| l| c| c c@{}}
\toprule
\multirow{2}{*}{Team} & \multirow{2}{*}{System} & \multirow{2}{*}{Ensem} & \multicolumn{2}{c}{EER (\%)} \\
& & & Eval & Test \\
\midrule
DFKI & EAT-L-SSLAM+BiCrossMamba & 5 & -- & \textbf{0.25} \\
AHU & EAT+AASIST & 5 & -- & \textbf{0.25} \\
CAU\cite{cau} & BEAT2AASIST & 3 & -- & 0.35 \\
CUC\cite{CUC} & SSLAM+FFN & 1 & 1.24 & 1.05 \\
HEU & BEATs+KNN & 1 & 40.91 & 2.96 \\
\midrule
Baseline 1\cite{esdd_eval} & BEATs+AASIST & 1 & 12.64 & 12.48 \\
Baseline 2\cite{esdd_eval} & AASIST & 1 & 15.72 & 15.40 \\
\bottomrule
\end{tabular}%
}
\vspace{-7mm}
\end{table}




\subsection{Track 1: ESDD in Unseen Generators}
\label{sec:track1}

Track~1 focuses on improving generalization to unseen generators. Here, we conclude the main strategies proposed by participants for addressing this challenge as follows.

\textbf{Model Designs:} Most teams improved generalization by combining strong pre-trained front-ends with more expressive back-end classifiers~\cite{cau,but,jaist}. Our Baseline~1 uses the Self-Supervised Learning (SSL) model (i.e., BEATs) to extract acoustic representations, and CAU further built on this direction. Specifically, they proposed BEAT2AASIST, which splits BEATs-derived features along frequency and channel dimensions and processes the features with two AASIST branches. Several top teams adopted newer SSL models, including EAT~\cite{eat} and SSLAM~\cite{sslam}, to obtain more transferable representations; DFKI additionally used attentive layer fusion to combine shallow and deep SSL features and employed BiCrossMamba-ST, a bidirectional selective state-space model for spectro-temporal modeling. Beyond these, CAU explored multiple top-$k$ transformer layer fusion strategies for classification to learn discriminative representations.

\textbf{Training Strategies:} Many teams improved robustness with data augmentation and task-specific objectives. AHU used targeted augmentation, including semantically-aligned data construction, MP3 compression, and loudness normalization. DFKI applied a ``Cut and Mix'' strategy and incorporated additional real data from AudioCaps~\cite{audiocaps}. CAU augmented training with vocoder-generated fake audio, while others used feature-domain augmentation to better handle unseen spectral distortions. To address class imbalance, CUC adopted a class-weighted training objective, and BIT combined LoRA~\cite{hu2022lora} fine-tuning, domain adversarial training~\cite{ganin2016domain}, and ArcFace~\cite{deng2019arcface} loss to improve generalization.


Several teams employed ensemble approaches for better performance. In summary, top-ranked teams combine SSL-based front-ends with AASIST-style back-ends, and ensemble-based systems outperform single-model approaches. The best performance is obtained by AHU, achieving an EER of 0.30\% on the test set.

\subsection{Track 2: Black-Box Low-Resource ESDD}

Initially, we expected Track 2 to be more challenging than Track 1, as models are required to generalize to an unseen generation framework (video-to-audio). However, the experimental results reveal that Track 2 generally presents lower EERs than Track 1, suggesting that the unseen generation framework does not introduce a significant additional challenge for current systems. The approaches in Track 2 largely follow the methods described in Section \ref{sec:track1}.

As shown in Table 2, ensemble-based systems dominate the top rankings, with the best-performing submissions achieving EERs as low as 0.25\%, while baseline systems exhibit substantially higher error rates, highlighting the effectiveness of advanced ensemble and SSL-based approaches in low-resource black-box settings.

\section{Conclusions and Future Scope}
In this paper, we introduce the ICASSP 2026 ESDD Challenge. The strong performance achieved by the participating teams demonstrates the effectiveness of current deepfake detection methods, even under unseen generators and low-resource conditions. At the same time, the results reveal remaining challenges in generalization and robustness across diverse sound types and generation settings. In future, we plan to further expand the dataset with more realistic and diverse deepfake scenarios and encourage the development of more robust and explainable detection approaches.

\balance
\bibliographystyle{IEEEbib}
\bibliography{strings,refs}

\end{document}